\documentstyle[PASJadd]{PASJ95}
\draft
\newcommand{\vol}{51}
\newcommand{\no}{5}
\newcommand{\lett}{}
\newcommand{\spage}{}
\newcommand{\rdate}{1999 January 8}
\newcommand{\adate}{1999 August 24}

\begin{document}
\setcounter{page}{0}

\title{Spectroscopic Study of the Vela-Shrapnel}

\author{Hiroshi {\sc Tsunemi} and Emi {\sc Miyata}\\
{\it  Department of Earth and Space Science, Graduate School of
Science, Osaka University,}\\ 
{\it 1-1 Machikaneyama-cho, Toyonaka, Osaka 560-0043 } \\
{\it E-mail(HT): tsunemi@ess.sci.osaka-u.ac.jp}\\
and \\
Bernd {\sc Aschenbach} \\
{\it Max-Planck-Institut f\"ur extraterrestrische Physik,
D-85740, Garching, Germany}}

\abst{Several shrapnels have been detected in the vicinity of Vela SNR
by the ROSAT all-sky survey.  We present here the spectral properties of
shrapnel `A' observed with the ASCA satellite. A prominent Si-K emission
line with relatively weak emission lines from other elements have been
detected, revealing that the relative abundance of Si is a few ten-times
higher than those of other elements.  Combining with the ROSAT PSPC
results, we obtained the electron temperature, $kT_{\rm e}$, to be $0.33 \pm
0.01$ keV.  The total mass of shrapnel `A' is estimated to be
$\sim 0.01 M_\odot$.  If it is an ejecta of a supernova explosion, the
interstellar matter (ISM) would be swept up in the leading edge while
the ejecta material would be peeled off in the trailing edge, which
should be confirmed by future observations.}

\kword{ISM: individual (Vela supernova remnant) ---
ISM: supernova remnants --- ISM: abundances --- X-Rays: ISM }

\maketitle
\thispagestyle{headings}

\section{Introduction}

The nucleosynthesis process inside a star generates high-$Z$ elements, and
supernova explosions spew them out into interstellar space.  Young
supernova remnants (SNRs), like Cassiopeia-A (Holt et al. 1994), Tycho
(Hwang, Gotthelf 1997) and Kepler's SNR (Kinugasa, Tsunemi 1999), show
clear X-ray emission lines from various heavy elements, indicating their
overabundance. These emission lines are mainly produced by ejecta,
enriched by heavy elements.

Middle-aged SNRs, like the Cygnus Loop, show X-ray emission lines
produced by swept-up interstellar matter (ISM).  The metal abundances
in the shell region are rather low (Miyata et al. 1994).  Since the
swept-up matter is much more massive than the ejecta, the apparent
abundances in these systems represent the ISM composition.  The Cygnus
Loop has a shell structure, showing that the major constituents are
swept-up matter in the shell region.  Whereas, Miyata et al. (1998)
detected a substantial amount of heavy elements in the center region of
the Loop.  The X-ray spectra show an overabundance by a factor of several
or more compared with the cosmic values.  They concluded that the heavy
elements were left in the core of the Cygnus Loop, probably a fossil
of the ejecta.  Therefore, matter enriched by heavy elements is left,
and still exists, even in the middle-aged SNR.

The kinds of heavy elements produced in a supernova depend on the type
of supernova explosion.  High-$Z$ elements, like A, Ca and Fe, are
generated dominantly in a Type-I supernova (Nomoto et al. 1984 ) while
low-$Z$ elements, like O, Ne, Mg and Si, are generated in a Type-II
supernova (Thielemann et al. 1996).  Comparing the relative metal
abundances with model calculations, Miyata et al.\ (1998) concluded
that the Cygnus Loop originated from a Type-II supernova, suggesting a
massive star as the progenitor.

Aschenbach et al.\ (1995, called AET hereafter) observed the entire Vela
SNR, which is also a typical middle-aged one, in the ROSAT all-sky
survey.  The SNR clearly shows a circular structure with a diameter of
$4.^{\!\!\circ}5$, probably consisting of swept-up ISM\@.  Furthermore, they
detected `shrapnels', boomerang structures (from `A' to `F'), outside
of the main shell.  The opening angles of the shrapnels suggest
supersonic motion in a tenuous matter.  These structures agree well with
the assumption that they originated in the center of the main shell,
which is very close to the Vela pulsar, 0833$-$45.

If the shrapnels are fossil material of the supernova explosion, we
expect that their spectra would show a metal-rich composition regardless
of the type of supernova.  There have been several fragments discovered
so far (AET).  Among them, fragment `A' has been selected for an ASCA
observation. This fragment is neither the brightest nor the largest one.
Since its apparent size is $4'\times 7'$, it is suitable for a study
with the ASCA SIS (Yamashita et al.\ 1997), which has a field of view
(FOV) of $11'$ square for one CCD chip.  AET found that there was a
relatively hot gas surrounding Vela SNR.  Therefore, the ambient
condition around Vela SNR might be different from that of the
standard SNR.  From these viewpoints, we selected fragment `A' in order
to observe the source and the background simultaneously.  The GIS
instrument (Ohashi et al. 1996) has a large circular FOV with a diameter
of $\sim 50'$ with a higher detection efficiency at higher energies,
while it has a less efficiency in a lower energy range with poorer energy
resolution than the SIS.  Therefore, the GIS is not very suitable to
determine the background spectrum for the SIS data.  This paper
describes the results of a combined analysis of the ASCA and ROSAT data.

\section{Observation and Results with ASCA}

We performed observations of fragment `A' during 1994 May 31 -- June
3, with the four-CCD mode at a high bit rate and with the two-CCD mode in
medium bit rate, resulting the effective exposure time of 91
ks.  Figure~\ref{GIS_map} shows the X-ray surface brightness map
obtained with two GIS sensors in a logarithmic gray scale, where the SIS
FOV is shown by squares.  There are two sources in the FOV: the left
one is a serendipitous source (CU Vel, a point source) and the right one
is shrapnel `A' (an extended source).

The SIS, functioning in four/two CCD modes, is pointed at $\alpha =
08^{\rm h} 57^{\rm m} 13^{\rm s}$, $\delta = - 41^{\circ}53^\prime
22^{\prime\prime}\ (2000)$.  This enables us to cover the major part of
fragment `A' in one CCD chip, while other chips cover both the tail
part and the surrounding region.  The observed region is divided into
two parts.  One is a circular region with a radius of $6\,'$ centered on
fragment `A'.  The other is its surrounding region excluding the
serendipitous source (CU Vel) in the east of fragment `A'.  This
configuration is shown in figure~\ref{GIS_map}.  The spectrum outside
the circular region is found to be statistically consistent with the
standard background, accumulated in the Lynx field and in the NEP field
(Gendreau et al.\ 1995).  Therefore, we subtracted the standard
background spectrum from the present data to obtain better statistics.
The count rate inside the circular region is $2.4\times 10^{-2}$ c
s$^{-1}$ /SIS and $1.6\times 10^{-2}$ c s$^{-1}$ /GIS, respectively.

We further divided the circular region into two parts: the northeast
half circle (leading part) and the southwest half circle (trailing
part).  Since spectral fits show no significant difference between the
two spectra in these regions, the spectral data in the circular region
were combined and treated together.

Figure~\ref{spectrum_SIS_all} shows the SIS spectrum with crosses.  The
spectrum is fairly smooth below 1~keV, whereas a strong emission line is
seen around 1.8 keV, which is considered to be the Si-K (Si {\small
XIII}) emission line.  There are two weak structures around 0.8 and 0.9
keV corresponding to the energies of Fe-L (Fe {\small XVII}) and Ne-K
(Ne {\small IX}) emission lines, respectively.  Various spectral models
were applied to examine the SIS data, as described below.  In the
spectral fits below, the interstellar absorption, $N_{\rm H}$, was fixed
at $3.5 \times10^{20}$ H atoms cm$^{-2}$, as determined by the ROSAT PSPC
(AET). This level of $N_{\rm H}$ is almost undetectable with the SIS.

\subsection{CIE Model with Cosmic Abundance}

The collisional ionization equilibrium (CIE) model with a
single-$kT_{\rm e}$ (Raymond, Smith 1977) was applied at first.  A
model with cosmic abundance (Anders, Grevesse 1989) gave
an unacceptable fit, as shown in table~1.  We then employed a model with
two different $kT_{\rm e}$ with cosmic abundance, since AET reported that
the PSPC spectrum could be well fitted with this model.  The results are
described in table~1.  We found that the model yielded statistically
unacceptable fits.

\subsection{CIE Model with Variable Abundance}

Next, we applied a single-$kT_{\rm e}$ CIE model with variable
abundances of elements.  In this fitting, the abundance of He was
fixed to the cosmic value, those of C and N are anchored to that of O,
and those of O, Ne, Mg, Si, Fe, and Ni were left as free
parameters. The best-fit model and parameters are shown in
figure~\ref{spectrum_SIS_all} and in table~1.  The major discrepancy
between the model and the data in this fit lies in the energy range
of 1.1 -- 1.2\,keV where the data are dominated by Fe-L line blends.

Then, we added an extra component with different $kT_{\rm e}$ and
cosmic abundances of elements.  This model gave us a moderately good
fit in the $\chi^2$ statistics.  However, a major deviation of the
model still exists in the energy range between 1.1--1.2\,keV,
indicating that the extra component is unable to fill up the gap
between the model and the data.  If we leave the abundances for both
components to be free, no meaningful parameters are obtained, because
the model has too many free parameters to adjust.

Through a fit with two temperature components, we found that the
spectrum in the energy range above 1.2\,keV could be well fitted with
the $kT_{\rm e} \sim 0.8$ keV component, while the data below $\sim 1$
keV could be well reproduced with the $kT_{\rm e}\sim 0.3$ keV
component. The obtained values of $kT_{\rm e}$ for these two components
are consistent with the PSPC results (AET)\@. The high-$kT_{\rm e}$
component is mainly responsible for both the continuum emission and the
Si emission line. The intensity ratio between the continuum and the Si
line constrains the value of $kT_{\rm e}$ when the cosmic abundance is
assumed.  If we do not fix the metal abundance, the variation of
$kT_{\rm e}$ would be inversely proportional to the relative abundance
of Si compared with those of other heavy elements.

\subsection{Fe-L Problem}

We applied various models, as described in the previous section, but
obtained no acceptable fits in the $\chi^2$ statistics.  A large
discrepancy in the energy range of 1.1 -- 1.2\,keV was found for all
models, particularly not in the GIS, but in the SIS data. The better
energy resolution seems to result in a larger deviation of the model.
This suggests that emission lines are involved in this part of the
spectrum.  Based on these results, we suspect that the poor spectral
fits result from improper knowledge of the Fe-L lines ($n=3 \rightarrow
2$), which mainly contribute to the spectrum in this energy range.  A
similar discrepancy was noticed in the SNR spectrum (Miyata et al.\
1998).  Since we need a multi-$kT_{\rm e}$ model to reproduce the
spectrum of the SNR, the disagreement between models and data may be
partly due to too simple modeling of the spectrum.  A similar problem
has also been reported in an analysis of the Centaurus cluster (Fabian
et al.\ 1994).  Liedahl et al. (1995) recalculated the emissivity of the
Fe-L line blend and found some deviation from the previous model.  They
claimed that the deviation was mainly caused by a problem in atomic
data.  Liedahl's calculation indicated that the emissivity of the $n=3
\rightarrow 2$ line became $30\%$ higher than that used in the current
plasma models, which could qualitatively solve our problem in the
spectral fits.  We employed the new model for thin thermal emission
(called VMEKAL in the XSPEC) and fit the data again (Mewe et
al. 1985). However, the new model did not significantly improve the fit.
The problem still seems to remain in the current models (Brickhouse et
al. 1995).  Since we could not access the revised atomic data, which
would solve the discrepancy, we decided to mask the energy range of 1.1
-- 1.2\,keV in the SIS spectral analysis, while no mask was applied for
the GIS data.

\subsection{Spectral Fits --- Ignoring Fe-L}

We applied single-$kT_{\rm e}$ and two-$kT_{\rm e}$ CIE models with the
cosmic abundance, but could not obtain a reasonable fit to the data.
Therefore, we fitted the SIS spectrum with a single-$kT_{\rm e}$ CIE
model with variable abundance excluding the data in the energy range of
1.1 -- 1.2\,keV\@.  In this fit, the $N_{\rm H}$ value was fixed to that
obtained by the ROSAT PSPC (AET)\@.  The best-fit curve and parameters
are shown in figure~\ref{spectrum_SIS_noFe} and in table~1. As shown in
the bottom panel of figure~\ref{spectrum_SIS_noFe}, the residuals show
random scatter, except for the energy range of 1.1 -- 1.2\,keV\@.  This
feature indicates that the model can be made statistically acceptable by
introducing certain systematic errors.  It should be noted that O is
strongly depleted in our fit. The $N_{\rm H}$ value employed here is
fairly low, and the SIS should be able to detect the O-line feature
easily in the spectrum.  Since the $N_{\rm H}$ value strongly correlates
with the O abundance in the spectral fit, the above conclusion
sensitively depends on the $N_{\rm H}$ value derived from the PSPC data.
If the O abundance was raised to the cosmic level, a substantially
higher value of $N_{\rm H}$ would be required in order to reproduce the
SIS spectrum. This is obviously inconsistent with the PSPC result.

The abundance values given in table~1 are based on the assumption that
He abundance is cosmic and C and N have equal abundance to O.  Even
though we have no effective way to confirm how adequate this assumption
is, we can claim that abundance ratio between O and Si is much smaller
than the cosmic value.  If the abundances of He, C, N, and O were assumed
to be cosmic, Si would have to be about several hundred times more
abundant than the cosmic level (Tsunemi, Miyata 1997).

\section{Combined Analysis with the ROSAT Data}

We obtained a consistent spectral fit between the SIS data and the GIS
data by ignoring the energy range of 1.1 -- 1.2 keV.  Since both
instruments are insensitive to the energy range below 0.4\,keV, we have
to fix the $N_{\rm H}$ value to the previously reported one (AET).  PSPC
is sensitive down to 0.1\,keV (Pfeffermann et al.\ 1987), while its
energy resolution is not sufficient to determine the metal abundances.
We checked the model by incorporating the PSPC data, which were obtained
with a pointing observation on 1992 May 31 -- June 1.  Data screening
was performed with $esas$ developed by Snowden (Snowden et al.\ 1994;
Snowden 1995).  The net exposure time was $\sim 8$\,ks.

We performed model fitting using all three data sets: the SIS, the GIS,
and the PSPC.  The model spectrum used is the CIE model described
before.  The best-fit curve and parameters are shown in
figure~\ref{spectrum_CIE} and in table~1.  The $N_{\rm H}$ value
obtained from our analysis is consistent with that obtained by the PSPC
alone (AET).  The metal abundances obtained from the three sets of data
are, however, a few-times larger than those obtained with the SIS data
only.  We notice that the relative abundances are similar to the
previous results on the SIS data only.  Since the PSPC is sensitive to
the low-energy region where the emission lines from C and N play a
dominant role, it might be inappropriate to assume that the abundances
of C and N are equal to that of O\@.  However, we could not practically
obtain meaningful values of the C and N abundances when we left them as
free parameters.  Based on these spectral fits, we can conclude that the
abundance ratio between O and Si is $8^{+6}_{-4}\times 10^{-2}$ times
the cosmic value, much smaller than the cosmic abundance ratio.

It is not surprising that the CIE model does not give us a satisfactory
fit.  We then employed a non-equilibrium ionization (NEI) model and
fitted all three data sets, simultaneously.  The only NEI code
practically available at present is the Masai model (Masai 1984; Masai
1994) which does not employ the revised atomic code (Liedahl et al.\
1995).  The best-fit curve and parameters are also shown in
figure~\ref{spectrum_NEI} and in table~1.  The best-fit ionization
timescale, $\log\tau$, is 11.4$\pm$0.2.  This indicates that the plasma
condition is relatively close to CIE, which is approximately achieved at
$\log\tau\sim 12$.  Therefore, the spectrum should be consistently
expressed with the CIE model.  The absolute abundance predicted by the
spectral fits depends on the employed model, and is strongly correlated
with other parameters, whereas we find that the abundance ratio between
O and Si, $6^{+4}_{-3}\times 10^{-2}$ for the NEI model, is relatively
robust.

\section{Discussion}

We have observed shrapnel `A' with ASCA.  The high-energy
resolution of the detector reveals relatively strong emission lines
around 1.8\,keV corresponding to the Si-K emission lines.  The other
interesting spectral feature is that almost no emission lines are seen
around 0.6 -- 0.7\,keV, the energy band for the O-K lines.  These results
suggest high Si and low O abundances.  In the spectral analysis, we
assumed the abundance ratio among C, N, and O to be equal to the cosmic
value, while the absolute abundance value was left as a free parameter.

Low abundances of C, N, and O are, thus, resulted due to a lack of O
emission lines.  Consequently H and He contributions dominate the
spectrum, further reducing the abundances of high-$Z$ elements.
Therefore, the absolute abundance values given in table 1 may not be
correct.  The absolute abundance of each element should be examined by
resolving all of the emission lines with high-resolution instruments in
the future.  However, the relative abundance derived here should be more
reliable, and it clearly shows that Si is extremely overabundant
compared with O.  This feature strongly supports the idea that
shrapnel `A' mainly contains supernova ejecta rather than the
ISM.

Strom et al. (1995) detected radio emission from the leading edge of
shrapnel `A', showing shock heating of the ambient medium
by the supersonic motion.  Shrapnel `A' is about
$5.^{\!\!\circ}2$ away from the center of the main shell, indicating
the actual distance $d$ to be $45.5\times(D/500\ {\rm pc})/
\sin\theta~{\rm pc}$, where $\theta$ is the angle between the line of sight
and the moving direction of shrapnel `A'.  The mean velocity,
$v_{\rm mean}$, is $4400\ (d/45.5\,{\rm pc}) (t/10000\,{\rm
yr})^{-1}$ km s$^{-1}$, where $t$ is the age of the SNR, whereas the
current velocity, $v_{\rm current}$, is $200\,{\rm km\ s^{-1}} (kT_{\rm
e}/0.3\,{\rm keV})^{0.5}$.  This indicates an order-of-magnitude
deceleration.

AET reported the detection of a tenuous plasma with a high temperature
in the vicinity of Vela SNR.  Based on their PSPC result and the
opening angle of the cone, the temperature of shrapnel `A' is expected
to be about 0.6 keV, which is two-times higher than that which we
obtained here.  Due to the difference in the energy resolving power
between the ASCA SIS and the ROSAT PSPC, we believe that the actual
temperature of shrapnel `A' is around 0.3 keV.

If shrapnel `A' is actually a moving ejecta, it would be slowed down by
the Ram pressure.  The deceleration is estimated to be $dv/dt =
-(\rho/\rho_0) (v^2/{\it l~f})$, where $v$, $\rho_0$, {\it l}, and {\it
f}\ are the velocity, the density, the length of shrapnel, and its
filling factor, respectively, while $\rho$ is the density of the ambient
matter.  If the shrapnel was injected during the supernova explosion
with an initial velocity of $\sim 10^4\,{\rm km \ s^{-1}}$ (Shigeyama et
al.\ 1994), it should have lost almost all kinetic energy, since $v_{\rm
current}$ is on the order of $\sim 10^2\,{\rm km\ s^{-1}}$.  If the
shrapnel has been keeping its shape since the explosion, we can
approximate $\rho d \sim \rho_0{\it l~f}$.  We can estimate $\rho_0$ to
be $2\times 10^{-2}$ H atoms cm$^{-3}$ based on the data in table~1
assuming that {\it f}\ is unity.  This supports the picture that the
explosion of Vela SN occurred in a bubble of hot tenuous gas.  The
temperature of the ambient medium is as high as that expected for an old
SNR, while the pressure is very close to the ISM level due to its low
density.

The total mass of shrapnel `A' is estimated to be $\sim
10^{-2} {\it f}^{0.5} M_{\odot}$.  Because of the relatively high abundance
of Si, it is likely that the shrapnel originated from the inner
part of the progenitor star.  If it is the case, the amount of H must
be relatively small, resulting in a decrease of the estimated mass.  A
two-dimensional hydrodynamic calculation of a supernova explosion
reveals the formation of a Rayleigh--Taylor instability and a convective
instability (Burrows et al. 1992).  Young et al.\ (1997) proposed
a double-supernova model for Vela SNR: one exploded about
150000 years ago, forming a tenuous high-temperature plasma, and the
other exploded some 10000 years ago.  They reported that the
collision of the secondary supernova shock with the primary ejecta
could produce fragmented ejecta with various abundance anomalies.

Assuming that shrapnel `A' is really a debris of the explosion of the
progenitor star, the leading part should have been contaminated by the
ISM and the trailing part should have lost some of the original material
by peeling.  The top part would consist of swept-up matter, while the
bottom part would show the raw composition of the shrapnel.  The
swept-up matter would be formed by two parts: the inner part is the
swept-up matter of the supernova ejecta while the outer part is the
swept-up ISM surrounding the SNR.  Therefore, we expect stratification
of three layers in the top part: the ISM, the ejecta and the original
matter.  The thickness of each layer would have an angular scale of
around $30^{\prime\prime}$, which is far beyond the imaging capability
of ASCA.  Such further evidence of the debris would be obtained by
observing the shrapnel with high spatial resolution.

\section{Conclusion}

We have observed shrapnel `A' in Vela SNR with ASCA.  The X-ray
spectrum shows a prominent Si-K emission line, and a spectral analysis
indicates an abnormal metal abundance.  The abundance ratio between O
and Si is only $4^{+3}_{-2}\times 10^{-2}$ times the cosmic value,
implying that a relative abundance of Si is a few tens of times higher
than that of O.  Combining with the ROSAT PSPC data, we still have a
relative abundance anomaly between Si and other elements.  We should
note that the relative abundances of heavy elements are consistent to be
the cosmic values, except for Si.

The mass of shrapnel `A' is $\sim10^{-2} {\it f}^{0.5} M_{\odot}$, which
can be a debris of the progenitor star of the supernova explosion.  The
current velocity is about 200\,km s$^{-1}$, while its initial velocity
is expected to be on the order of $10^4$\,km s$^{-1}$.  It is still a
question how the shrapnel was formed and survived through passage of the
SNR shell.  A spatially resolved spectroscopic study in the future
should confirm its nature, whether or not it really comes from the
progenitor star.

\vspace{1pc}\par The authors thank all members of the ASCA
team. E.M. acknowledges the hospitality of J. Tr\"umper and the X-ray
astronomy group in MPE. This research has made use of data obtained
from the High Energy Astrophysics Science Archive Research Center
Online Service, provided by NASA/Goddard Space Flight Center.

\clearpage
\section*{References}
\re
Anders E., Grevesse, N. 1989, Geochim. Cosmochim. Acta 53, 197
\re
Aschenbach B., Egger R., Tr\"umper J. 1995, Nature 373, 587 (AET)
\re
Burrows A., Fryxell B. A., 1992, Science 258, 430
\re
Brickhouse N., Edgar R., Kaastra J., Kallman T., Liedahl D., Masai K.,
Monsignori-Fossi B. et al. 1995, Legacy 6, 4
\re
Fabian A.C., Arnaud K.A., Bautz M.W., Tawara Y. 1994, ApJ 436, L63
\re
Gendreau K.C., Mushotzky R., Fabian A.C., Holt S.S., Kii T., Serlemitsos
P.J., Ogasaka Y., Tanaka Y. et al. 1995, PASJ 47, L5
\re
Holt S.S., Gotthelf E.V., Tsunemi H., Negoro H. 1994, PASJ 46, L151
\re
Hwang U., Gotthelf E.V. 1997, ApJ 475, 665
\re
Kinugasa K., Tsunemi H. 1999, PASJ 51, 239
\re
Liedahl D.A., Osterheld A.L., Goldstein W.H. 1995, ApJ, 438, L115
\re
Masai K. 1984, Ap\&SS 98, 367
\re
Masai K. 1994, ApJ 437, 770
\re
Mewe R., Gronenschild E.H.B.M., van den Oord G.H.J. 1985, A\&AS 62, 197
\re
Miyata E., Tsunemi H., Kohmura T., Suzuki S., Kumagai S. 1998, PASJ 50, 257
\re
Miyata E., Tsunemi H., Pisarski R., Kissel S.E. 1994, PASJ 46, L101
\re
Nomoto K., Thielemann F.-K., Yokoi K. 1984, ApJ 286, 644
\re
Ohashi T., Ebisawa K., Fukazawa Y., Hiyoshi K., Horii M.,
    Ikebe Y., Ikeda H., Inoue H. et al. 1996, PASJ 48, 157
\re
Pfeffermann E., et al. 1987, Proc. SPIE 733, 519
\re
Raymond J.C., Smith B.W. 1977, ApJS 35, 419
\re
Shigeyama T., Suzuki T., Kumagai S., Nomoto K., Saio H., Yamaoka H.
1994, ApJ 420, 341
\re
Snowden S.L. 1995, Cookbook for Analysis Procedures for ROSAT XRT/PSPC Observations of Extended Objects and the Diffuse Background
\re
Snowden S.L., McCammon D., Burrows D.N., Mendenhall J.A. 1994, ApJ 424, 714
\re
Strom R., Johnson H. M., Verbunt F., Aschenbach B. 1995, Nature 373, 590
\re
Thielemann F.-K., Nomoto K., Hashimoto M. 1996, ApJ 460, 408
\re
Tsunemi H., Miyata E., 1997, in Thermonuclear Supernovae, ed
P. Puiz-Lapuente, R. Canal, J. Isern (Kluwer Academic Publishers, Netherlands)
 p561
\re
Yamashita Y., Dotani T., Bautz M., Crew G., Ezuka H., Gendreau
K., Kotani T., Mitsuda K. et al. 1997, IEEE Trans. Nucl. Sci. 44, 847
\re
Young T. R., Shigeyama T., Suzuki T. 1997, in X-Ray Imaging and
Spectroscopy of Cosmic Hot Plasmas, ed F. Makino, K. Mitsuda (University
Academy Press, Tokyo) p405

\clearpage

\begin{figure}
\caption{X-ray intensity map obtained with the GIS shown in a
logarithmic gray scale.  The FOV of the SIS is superposed with squares,
while the circle shows the region from which we extracted the spectrum.
Two sources are seen: the left one is a serendipitous source (CU Vel, a
point source) and the right one is shrapnel `A' (an extended
source).}  \label{GIS_map}
\end{figure}

\begin{figure}
\caption{X-ray spectrum obtained with the SIS. The crosses show the data
points with $\pm$ 1$\sigma$ errors. The solid line shows the best-fit
curve of the single-$kT_{\rm e}$ CIE model with variable abundance and
the lower panel shows the residuals of the fit.}
\label{spectrum_SIS_all}
\end{figure}

\begin{figure}
\caption{Same as figure 2, but ignored between 1.1
-- 1.2\,keV (shown as dotted crosses in lower panel).} 
\label{spectrum_SIS_noFe}
\end{figure}

\begin{figure}
\caption{Combine fitting for the SIS, the GIS, and the PSPC spectra using the CIE model.}
\label{spectrum_CIE}
\end{figure}

\begin{figure}
\caption{Same as figure 4, but using the NEI model.}
\label{spectrum_NEI}	
\end{figure}

\clearpage

\begin{table}[htbp]
  \caption{{\sc Fitting Results}}
  \label{table:fit}

  {\footnotesize
  \begin{tabular}{c c c c c c c c c c c c c c}
    \hline\hline
    Model & Red. $\chi^2$ (d.o.f.) &
	kTe & C,N,O & Ne & Mg & Si & Fe & EM$^{\rm b}$ &
		Density$^{\rm c}$\\
	& & [keV]  & & & & & & [pc$^3$ cm$^{-6}$] &
		 [cm$^{-3}$]\\
    \hline
    1-kTe cosmic & 13 (83) & 0.40 &
	1$^{\rm a}$ & 1$^{\rm a}$ & 1$^{\rm a}$ & 1$^{\rm a}$ & 1$^{\rm a}$ &
	0.10 & 0.71 \\
    2-kTe cosmic & 2.7 (81) & 0.21/0.78 &
	1$^{\rm a}$ & 1$^{\rm a}$ & 1$^{\rm a}$ & 1$^{\rm a}$ & 1$^{\rm a}$ & 
	0.10/8.4$\times 10^{-2}$ & 0.71/0.64 \\
    1-kTe variable & 1.9 (77) & 0.32 &
	2$\times 10^{-2}$ & 8$\times 10^{-2}$ &
	2$\times 10^{-2}$ & 0.5 & 2$\times 10^{-2}$ & 
	3.2 & 3.9 \\
	\hline
	\multicolumn{2}{c}{1-kTe with variable model} \\
	\hline
	SIS$^{\rm d}$ & 1.3 (72) & 0.31 $\pm$ 0.02 &
		$3^{+3}_{-2} \times 10^{-2}$ &	
		$0.15^{+0.09}_{-0.05}$ &			
		$0.11^{+0.1}_{-0.06}$ & 			
		$1.0^{+0.5}_{-0.3}$ &			
		$4^{+2}_{1} \times 10^{-2}$ &	
		$1.9^{+0.8}_{-0.5}$ &
		$3.1^{+0.7}_{-0.5}$ \\
	GIS & 1.4 (108) & $ 0.28^{+0.01}_{-0.03}$ & & & & & &
		$2.4^{+1.1}_{-0.3}$ & $3.5^{+0.6}_{-0.2}$ \\
	PSPC$^{\rm e}$ & 2.0 (16) &
		$0.26^{+0.02}_{-0.01}$ & & & & & &
		$2.3^{+0.5}_{-0.4}$ & $3.3^{+0.4}_{-0.3}$ \\
    \hline
	Combine$^{\rm f}$ & 1.5 (199) & 0.30 $\pm$ 0.02 &
		$5^{+4}_{-2} \times 10^{-2}$ &
		0.2 $\pm$ 0.1 &
		0.2 $\pm$ 0.1 &
		$1.5^{+0.8}_{-0.5}$ &
		$7^{+4}_{-2} \times 10^{-2}$ &
		1.0 $\pm$ 0.4 &
		2.4 $\pm$ 0.5\\
    \hline
  \end{tabular}\\

   }{\small
	{\sc Note} --- The quoted errors are at 90~\% confidence
		level.\\
	$^{\rm a}$ Fixed to unity. \\
	$^{\rm b}$ EM (emission measure) is defined as
		$(\frac{D}{500})^2 {\rm n_e
		n_{\scriptscriptstyle H} V}$ [pc$^3$ cm$^{-6}$].
		Here, $D$ is distance [pc], V is volume [pc$^3$],
		${\rm n_e}$ is electron number density [cm$^{-3}$], and
		${\rm n_{\scriptscriptstyle H}}$ is hydrogen number
		density [cm$^{-3}$].\\
	$^{\rm c}$ Assuming the spherical symmetry and uniformity of
			the source and radius to be
			$2.5'$ $\approx$ 0.36~pc.\\
	$^{\rm d}$ Ignore 1.1 -- 1.2~keV range when fitting.\\
	$^{\rm e}$ Obtained $N_{\rm H}$ was (5 $\pm$ 1) $\times 10^{20}$
		H atoms cm$^{-2}$.\\
	$^{\rm f}$ Obtained $N_{\rm H}$ was (3.5 $\pm$ 0.8) $\times 10^{20}$
		H atoms cm$^{-2}$.
}

\end{table}

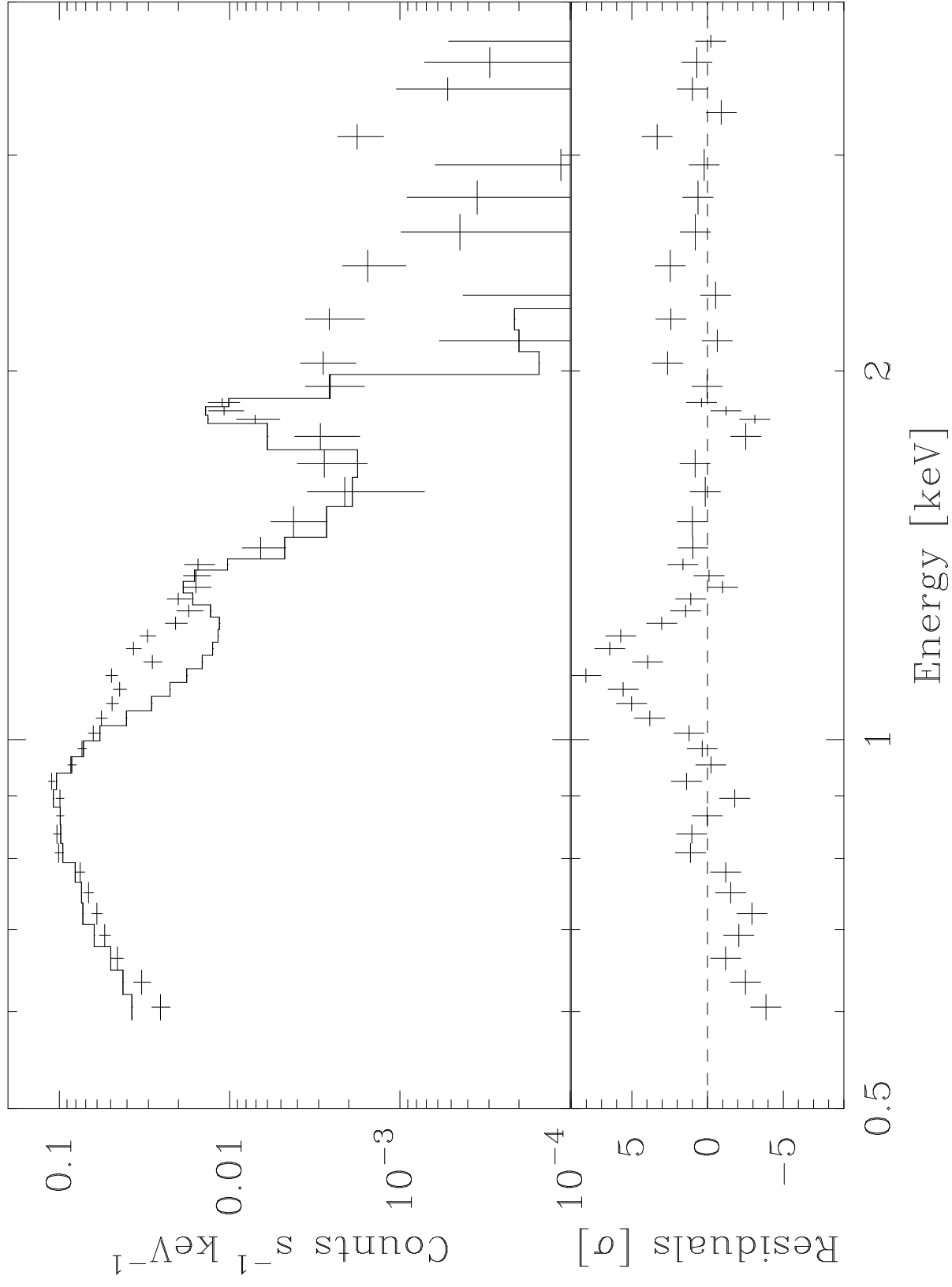
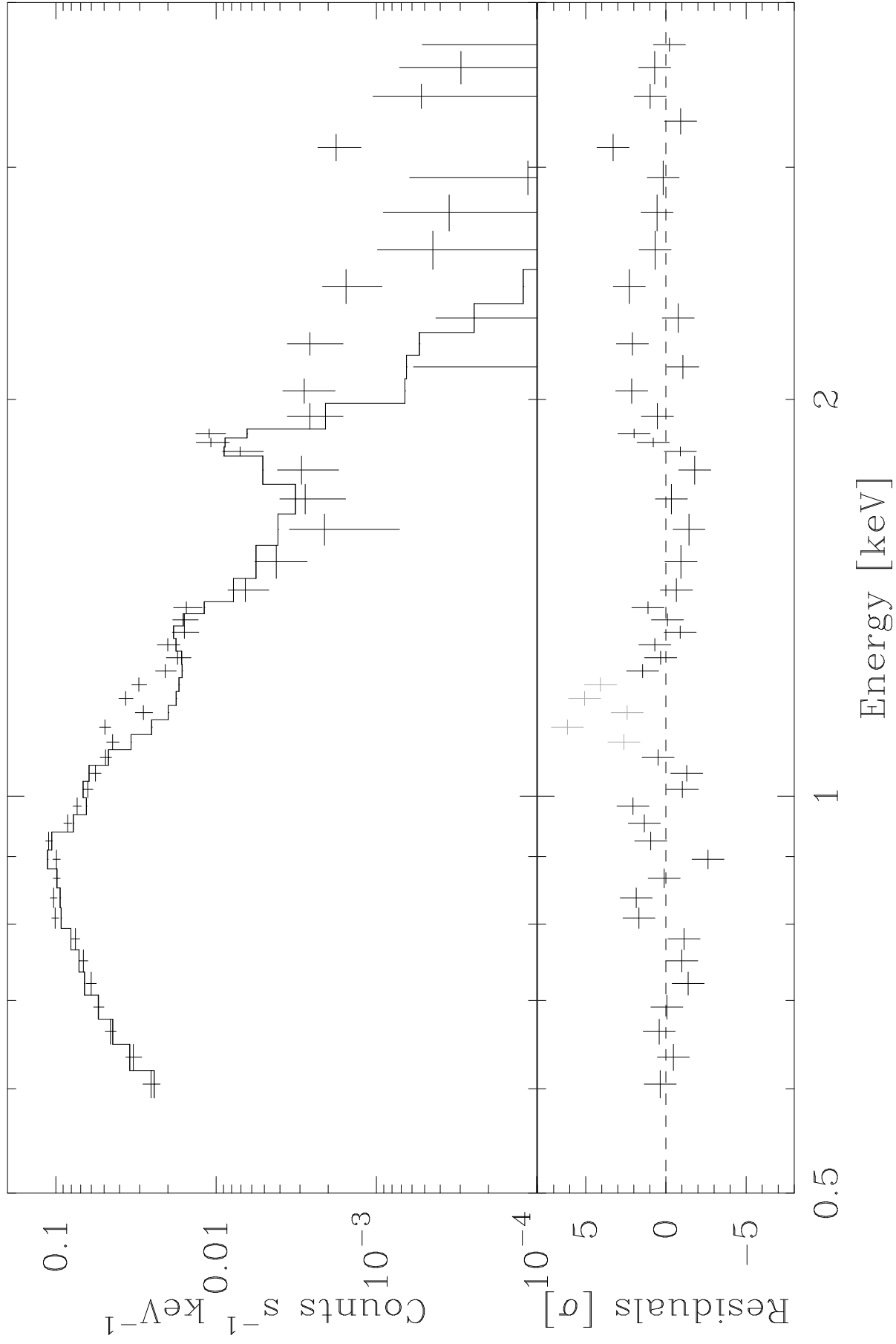
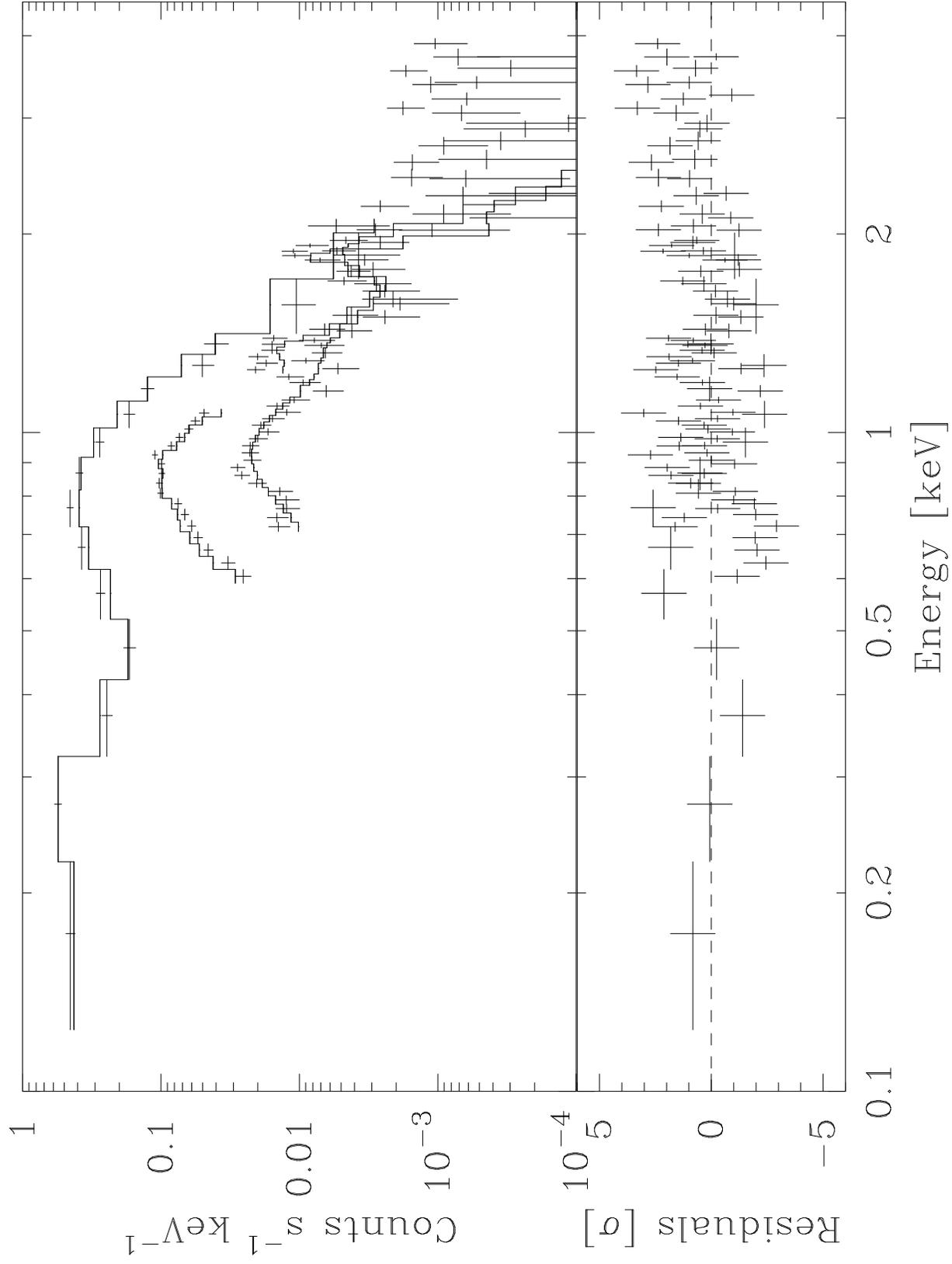
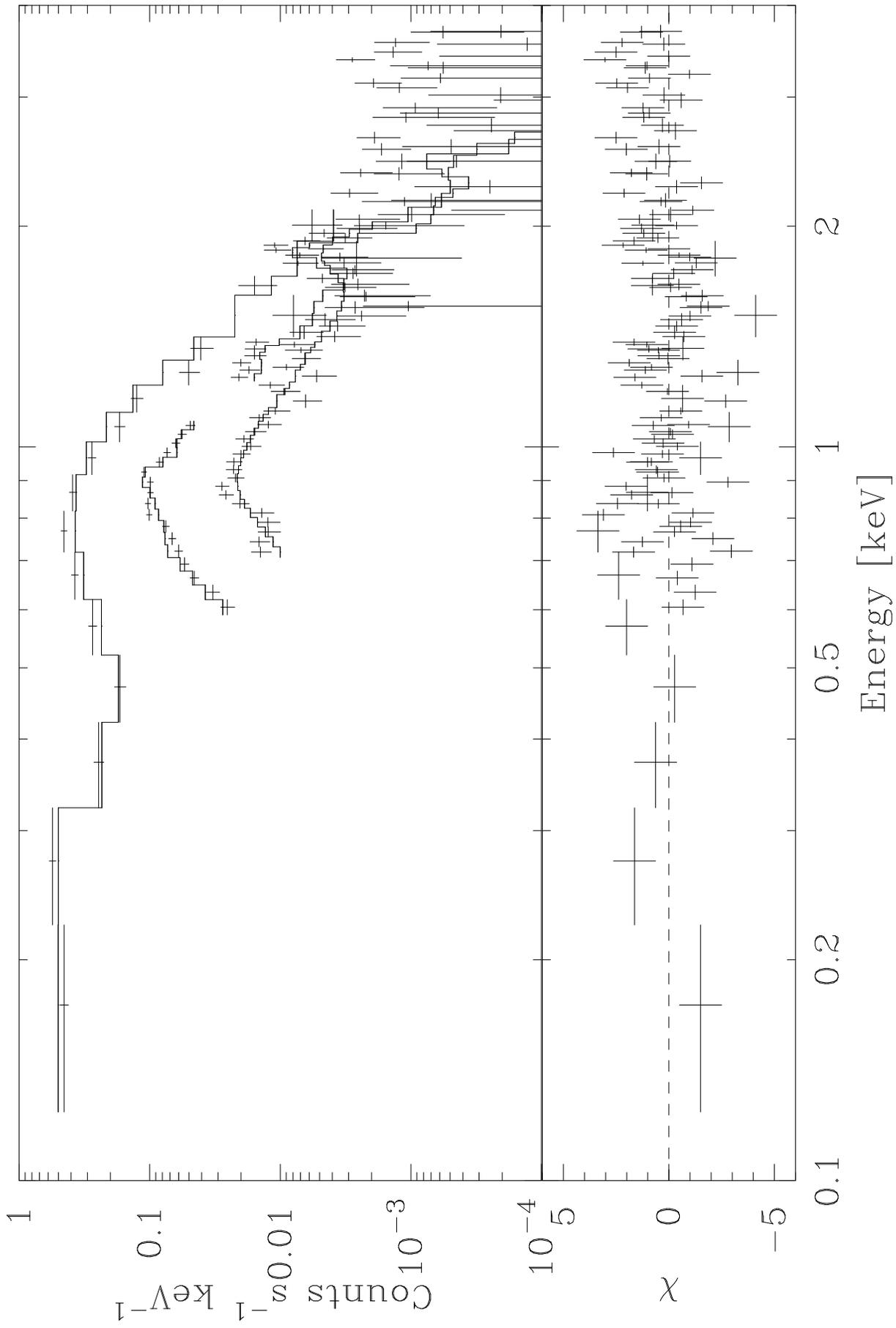
\end{document}